\documentclass[ALICE,manyauthors]{cernphprep}
\usepackage[comma,square,numbers,sort&compress]{natbib}
\usepackage{hyperref}
\usepackage{lineno}
\usepackage{xspace}

\usepackage{xcolor}
\usepackage[normalem]{ulem}
\usepackage{booktabs}
\usepackage{bm}
\usepackage[T1]{fontenc}
\usepackage{orcidlink}

\newcommand{\pp}           {pp\xspace}

\newcommand{\PbPb}         {\mbox{Pb--Pb}\xspace}

\newcommand{\snn}          {\ensuremath{\sqrt{s_{\mathrm{NN}}}}\xspace}

\newcommand{\nineH}        {$\sqrt{s}~=~0.9$~Te\kern-.1emV\xspace}
\newcommand{\seven}        {$\sqrt{s}~=~7$~Te\kern-.1emV\xspace}
\newcommand{\twoH}         {$\sqrt{s}~=~0.2$~Te\kern-.1emV\xspace}
\newcommand{\twosevensix}  {$\sqrt{s}~=~2.76$~Te\kern-.1emV\xspace}
\newcommand{\five}         {$\sqrt{s}~=~5.02$~Te\kern-.1emV\xspace}

\newcommand{\twosevensixnn}{$\sqrt{s_{\mathrm{NN}}}~=~2.76$~Te\kern-.1emV\xspace}
\newcommand{\fivenn}       {$\sqrt{s_{\mathrm{NN}}}~=~5.02$~Te\kern-.1emV\xspace}

\newcommand{\GeVc}         {Ge\kern-.1emV/$c$\xspace}
\newcommand{\MeVc}         {Me\kern-.1emV/$c$\xspace}
\newcommand{\TeV}          {Te\kern-.1emV\xspace}
\newcommand{\GeV}          {Ge\kern-.1emV\xspace}
\newcommand{\MeV}          {Me\kern-.1emV\xspace}
\newcommand{\GeVmass}      {Ge\kern-.2emV/$c^2$\xspace}
\newcommand{\MeVmass}      {Me\kern-.2emV/$c^2$\xspace}

\newcommand{\Lint}{\ensuremath{\mathcal{L}_\mathrm{int}}\xspace}
\newcommand{\invnb}{\ensuremath{\mathrm{nb}^{-1}}\xspace}

\newcommand{\moliere}{Moli\`ere}
\newcommand{\hybridWithMoliere}{Hybrid/\moliere}
\newcommand{\hybridWithoutMoliere}{Hybrid/no\moliere}

\newcommand{\pT}{\ensuremath{p_{\text{T}}}}
\newcommand{\pTJet}{\ensuremath{p_{\text{T,jet}}}}
\newcommand{\pTJetCh}{\ensuremath{p_{\text{T,ch jet}}}}
\newcommand{\kT}{\ensuremath{k_{\text{T}}}}
\newcommand{\kTg}{\ensuremath{k_{\text{T,g}}}}
\newcommand{\kTgMeas}{\ensuremath{k^{\text{meas.}}_{\text{T,g}}}}

\newcommand{\deltaR}{\ensuremath{\Delta R}}
\newcommand{\zcut}[1]{\ensuremath{z_{\text{cut}} = #1}}
\newcommand{\sdZ}[1]{soft drop \ensuremath{z_{\text{cut}} = #1}}
\newcommand{\dygA}[1]{dynamical grooming \ensuremath{a = #1}}

\newcommand{\sdZA}[1]{SD (\ensuremath{z_{\text{cut}} = #1})}
\newcommand{\dygAA}[1]{DyG (\ensuremath{a = #1})}

\begin{document}

%%%%%%%%%%%%%%%  Title page %%%%%%%%%%%%%%%%%%%%%%%%
\begin{titlepage}
% the dates below correspond to CERN approval
% please don't touch: EB chairs will take care
\PHyear{2024}       % required, will be obtained from CERN
\PHnumber{238}      % required, will be obtained from CERN
\PHdate{16 September}  % required, will be obtained from CERN
%%%%%%%%%%%%%%%%%%%%%%%%%%%%%%%%%%%%%%%%%%%%%%%%%%%%

%%% Put your own title + short title here:
\title{Search for quasi-particle scattering in the quark--gluon plasma with jet splittings in pp and Pb$-$Pb collisions at $\mathbf{\sqrt{\textit{s}_{\textup{NN}}} = 5.02}$ TeV}   % appears on left page headers
\ShortTitle{Search for quasi-particle scattering in the QGP with jet splittings}

%%% Do not change the next lines
\Collaboration{ALICE Collaboration\thanks{See Appendix~\ref{app:collab} for the list of collaboration members}}
\ShortAuthor{ALICE Collaboration} % appears on right page headers, do not change

\begin{abstract}
The ALICE Collaboration reports measurements of the large relative transverse momentum ($k_{\text{T}}$) component of jet substructure in pp and Pb$-$Pb collisions at center-of-mass energy per nucleon pair $\sqrt{s_{\text{NN}}}=5.02$ TeV.
Enhancement in the yield of such large-$k_{\text{T}}$ emissions in head-on Pb$-$Pb collisions is predicted to arise from partonic scattering with quasi-particles of the quark--gluon plasma.
The analysis utilizes charged-particle jets reconstructed by the anti-$k_{\text{T}}$ algorithm with resolution parameter $R=0.2$ in the transverse-momentum interval $60 < p_{\text{T,ch jet}} < 80$\:$\text{GeV}/c$.
The soft drop and dynamical grooming algorithms are used to identify high transverse momentum splittings in the jet shower.
Comparison of measurements in Pb$-$Pb and pp collisions shows medium-induced narrowing, corresponding to yield suppression of high-$k_{\text{T}}$ splittings, in contrast to the expectation of yield enhancement due to quasi-particle scattering.
The measurements are compared to theoretical model calculations incorporating jet modification due to jet-medium interactions ("jet quenching"), both with and without quasi-particle scattering effects.
These measurements provide new insight into the underlying mechanisms and theoretical modeling of jet quenching.
\end{abstract}
\vspace{1.5cm}
\begin{center}
\emph{This publication is dedicated to the memory of our colleague\\ Dr.\ Roy Crawford Lemmon who recently passed away}
\end{center}

\end{titlepage}

\setcounter{page}{2} %please do not remove this line

%%%%%%%%%%%%%%%%%%%%%%%%%%%%%%%%
% begin main text
%%%%%%%%%%%%%%%%%%%%%%%%%%%%%%%%

Strongly-interacting matter at high temperature and density forms a deconfined state known as the quark--gluon plasma (QGP)~\cite{Collins:1974ky, Shuryak:1977ut}.
The QGP filled the universe a few microseconds after the Big Bang, and it is generated and studied today using high-energy nuclear collisions at the Relativistic Heavy-ion Collider (RHIC) and the Large Hadron Collider (LHC)~\cite{STAR:2005gfr,PHENIX:2004vcz,ALICE:2022wpn,CMS:2024krd}.
RHIC and LHC measurements, together with theoretical studies, have revealed that the QGP exhibits marked emergent features; it flows with low specific viscosity and is opaque to the passage of energetic partons~\cite{Jacak:2012dx,Busza:2018rrf,Harris:2023tti}.
While static properties of the QGP can be calculated numerically using quantum chromodynamics (QCD) implemented on the lattice~\cite{Ratti:2018ksb,Guenther:2020jwe}, the microscopic mechanisms underlying these dynamical phenomena remain poorly understood.

In high-energy hadronic collisions, partons from the projectiles may scatter with high momentum transfer.
Such scattered partons are initially highly virtual and evolve by radiating secondary partons that hadronize in a collimated spray of particles which is observable experimentally, denoted a ``jet.''
The jet shower has rich internal substructure that is observable in correlations between its constituent particles.
In \pp{} collisions, jet production cross sections and substructure have been measured extensively~\cite{ALICE:2013yva,ATLAS:2012am,CMS:2021yzl,ATLAS:2020bbn,ALICE:2022hyz}.
Theoretical calculations at high perturbative order (pQCD) agree well with the measurements, providing stringent tests of QCD~\cite{STAR:2006opb, STAR:2016kpm, ALICE:2013yva, ATLAS:2014riz, CMS:2016lna, ATLAS:2020bbn}.
Charged-particle jets are especially well suited for jet substructure measurement, due to the excellent angular resolution of charged-particle tracking of current detectors. Their properties are calculable in pQCD using non-perturbative track functions~\cite{Lee:2023xzv}.

In collisions of heavy nuclei, jets are generated together with the QGP and interact with it.
The interaction of a jet and the QGP, known as jet quenching, generates several observable phenomena: jet yield suppression, jet deflection, and modification of jet substructure~\cite{Cunqueiro:2021wls, Apolinario:2022vzg}.
Jet quenching has been measured extensively at RHIC and the LHC with multiple observables~\cite{STAR:2020xiv,ALICE:2023waz,ATLAS:2018gwx,CMS:2021vui, ALICE:2015mdb,STAR:2017hhs, ATLAS:2022vii,ALargeIonColliderExperiment:2021mqf,CMS:2018fof,ALICE:2017nij,ATLAS:2023hso}, providing unique probes of the structure and dynamics of the QGP~\cite{JET:2013cls,JETSCAPE:2021ehl,JETSCAPE:2024cqe,Ke:2020clc}.

Jet scattering in the QGP may elucidate the microscopic structure of the medium and the nature of its constituent quasi-particles, in analogy to Rutherford scattering which revealed the atomic nucleus~\cite{Rutherford:1911zz}.
Rare, single-hard secondary scattering of a jet off of medium constituents would result in a Coulomb-like power-law tail in the emission spectrum (\moliere{} scattering)~\cite{DEramo:2012uzl,DEramo:2018eoy,Barata:2020rdn,Cunqueiro:2023vxl}.

Two experimental approaches have been proposed to search for direct evidence of jet--QGP scattering: in-medium deflection of the entire jet, which modifies the acoplanarity distribution~\cite{ALICE:2015mdb,STAR:2017hhs,ALICE:2023qve}, and deflection of a component of the jet shower, which modifies jet internal structure~\cite{DEramo:2012uzl,DEramo:2018eoy}.
While significant medium-induced modification to acoplanarity distributions has recently been reported, the dependence of the modification on jet resolution parameter $R$ indicates that it is due predominantly to medium response and energy redistribution, which mask the \moliere{}-scattering signal~\cite{ALICE:2023jye, ALICE:2023qve,STAR:2023ksv, STAR:2023pal, CMS:2017ehl}.
For substructure observables, the in-medium scattering of jet shower components off of QGP quasi-particles corresponds to a relative increase in the rate of high transverse momentum splittings (high $\kT{}$) in the jet shower~\cite{DEramo:2012uzl,DEramo:2018eoy}; to date, no search for \moliere{} scattering in the QGP based on jet substructure has been reported.

This Letter reports the first measurement of medium-induced modification of jet substructure using high-$\kT{}$ splittings to search for quasi-particle scattering in the quark--gluon plasma.
The data were recorded by the ALICE Collaboration during LHC Run 2 for \pp{} collisions and for central and semicentral \PbPb{} collisions at center-of-mass energy per nucleon pair $\snn{}$ = 5.02 TeV.
Charged-particle jets are reconstructed using the anti-$\kT{}$ algorithm~\cite{Cacciari:2008gn,Cacciari:2008gp} with resolution parameter $R$ = 0.2 in the $60 < \pTJetCh{} < 80$ \GeVc{} jet transverse momentum interval.
Background from the underlying event is corrected using constituent subtraction~\cite{Berta:2014eza}.
Jet substructure is measured using the soft drop~\cite{Larkoski:2014wba} and dynamical grooming~\cite{Mehtar-Tani:2019rrk} algorithms.
Theoretical calculations incorporating jet quenching are used to assess the effects of in-medium \moliere{} scattering, and are compared to the data.

The ALICE detector and its performance are described in Refs.~\cite{ALICE:2008ngc,ALICE:2014sbx,ALICE:2022wpn}.
Event triggering and event characterization utilize the V0A and V0C scintillator detectors, which cover pseudorapidity intervals $2.8 < \eta < 5.1$ and $-3.7 < \eta < -1.7$, respectively~\cite{ALICE:2013axi},
and the Zero Degree Calorimeters located $\pm112.5$ m along the beam axis from the nominal interaction point of the beams.
Charged-particle tracking is carried out using signals from the Inner Tracking System (ITS)~\cite{ALICE:2010tia} and Time Projection Chamber (TPC)~\cite{Alme:2010ke} (and Transition Radiation Detector (TRD)~\cite{ALICE:2017ymw} in \pp{} collisions). The tracking acceptance covers $|\eta| < 0.9$ over the full azimuth.

The dataset, triggering, and event selection for \pp{} and \PbPb{} collisions at \fivenn{} are identical to those of the analysis described in Ref.~\cite{ALargeIonColliderExperiment:2021mqf}.
Data were recorded with a minimum bias trigger and an event-activity (EA) trigger based on the V0 signal.
For the \PbPb{} analysis, events are selected within the 0--10\% (``central'') and 30--50\% (``semicentral'') decile bins in EA.
After event selection the \pp{} collision dataset comprises 870 million events, corresponding to integrated luminosity $\Lint=18.0\pm0.4$ \invnb~\cite{ALICE-PUBLIC-2018-014}, while the \PbPb{} dataset comprises 92 million central and 90 million semicentral events, corresponding to $\Lint=0.12$ and $0.06$ \invnb, respectively, with approximately 3\% uncertainty~\cite{ALICE:2022xir}.

Charged-particle tracks accepted for the analysis have $\pT > 0.15$ \GeVc{}.
Due to the spatially non-uniform ITS response, which also varied in time during the data-taking period, different classes of tracks are combined to create a population of charged-particle tracks with approximately uniform azimuthal acceptance~\cite{ALICE:2012nbx}.
Accepted tracks have momentum resolution $\sigma(\pT{})/\pT{}\sim1\%$ for track $\pT{} = 1$~\GeVc{} and 4\% for $\pT{} = 50$~\GeVc{}.

Both the \pp{} and \PbPb{} datasets are corrected using events simulated by the PYTHIA 8 generator~\cite{Sjostrand:2014zea} with the Monash 2013 tune~\cite{Skands:2014pea}, whose output is denoted ``particle level.''
These events are then passed through a detailed model of the ALICE apparatus using GEANT 3~\cite{geant3ref2} to simulate detector effects, and reconstructed with the same algorithms utilized for the event, track, and jet reconstruction of the data; such events are denoted ``detector level.''
For \PbPb{} collisions, detector-level events are additionally embedded into real \PbPb{} data to include the effect of background from the underlying event, and are referred to as ``reconstructed level''.

Measurements in \pp{} collisions are compared with particle-level calculations using  the \pp{}-collision mode of JEWEL~\cite{Zapp:2012ak,Zapp:2013vla} and the Hybrid model~\cite{Casalderrey-Solana:2014bpa}, which are based on PYTHIA 6~\cite{Sjostrand:2006za} and PYTHIA 8, respectively,
and the JETSCAPE PP19 tune~\cite{JETSCAPE:2019udz} based on the vacuum mode of the MATTER model~\cite{Cao:2017qpx}.
For \PbPb{} collisions, the data are compared to calculations incorporating jet quenching.
JETSCAPE (v3.5 AA22 tune~\cite{Putschke:2019yrg,JETSCAPE:2022jer,JETSCAPE:2023hqn}) provides a multi-stage calculation of medium-modified parton showers with virtuality-dependent interactions, based on MATTER and LBT~\cite{He:2015pra}.
The Hybrid Model~\cite{DEramo:2018eoy, Hulcher:2017cpt} incorporates AdS-CFT-inspired partonic energy loss and a model of medium response (``wake'').
JEWEL is based on the BDMPS energy loss formalism~\cite{Baier:1996kr} within a medium modeled by Bjorken expansion, with an option to explicitly account for the distribution of partons in the QGP that recoil from the jet-QGP interaction.
JETSCAPE and JEWEL include \moliere{} scattering, while the Hybrid Model can disable elastic $2\rightarrow{}2$ \moliere{} scattering to isolate such scattering effects.

The large underlying-event (UE) background in measured data is characterized via the $\pT{}$ density $\rho$ and mass density $\rho_{\text{m}}$~\cite{ALICE:2012nbx,ALICE:2017nij}.
This background is suppressed using event-by-event constituent subtraction~\cite{Berta:2014eza} with maximum subtraction distance parameter $R_{\text{max}}$ = 0.1, and $\pT{}$-weight $\alpha$ = 0, which is applied prior to jet finding.
No UE subtraction is applied in \pp{} collisions.
Jet clustering implemented in FastJet 3.3.3~\cite{Cacciari:2011ma} is carried out twice in each event: first applying the anti-$\kT{}$ clustering algorithm~\cite{Cacciari:2008gn,Cacciari:2008gp} with charged-particle tracks to reconstruct jet candidates, and then using the Cambridge--Aachen (CA) clustering algorithm~\cite{Dokshitzer:1997in} with jet constituents from selected jet candidates to examine the jet substructure.

The anti-$\kT{}$ clustering utilizes jet resolution parameter $R=0.2$ and the $E$-recombination scheme~\cite{Cacciari:2008gn}. The value $R=0.2$ was chosen as a balance between sensitivity to medium effects and measurement precision in the complex heavy-ion collision environment. The pion mass is assumed for all tracks.
Charged-particle jet candidates are required to have $40 \leq \pTJetCh{} < 120$ \GeVc{}, jet area $A_{\text{jet}} > 0.6\,\pi\,R^{2}$~\cite{ALICE:2013dpt}, and jet centroid $|\eta_{\text{jet}}| < 0.9 - R$, with the minimum $\pTJetCh{}$ and area requirements applied to remove jet candidates arising from uncorrelated background.
Jet candidates containing a constituent with $\pT > 100$ \GeVc{} are rejected due to reduced momentum resolution in this region.
After background subtraction, the area and constituent $\pT{}$ requirements remove fewer than 1\% of all jets.

Jet grooming algorithms~\cite{Chien:2016led, Mehtar-Tani:2016aco} are used to remove soft, wide-angle radiation by unwinding the CA recombination tree, allowing the identification of hard splittings.
At each recombination step the algorithm evaluates a specified grooming condition based on a ``hardness measure,'' either accepting and recording the splitting properties, or rejecting the splitting and following the harder subjet to the next splitting.
We utilize two different grooming algorithms which select splittings in distinct but partially overlapping regions of phase space, which may probe \moliere{} scattering in different ways.

For soft drop grooming (``SD'')~\cite{Larkoski:2014wba}, the hardness measure is the fraction of the momentum carried in the softer subjet, $z$, and the grooming condition is $z > z_{\text{cut}} (\Delta R/R)^{\beta}$, where $\Delta R$ is the angle between the subjets, and $z_{\text{cut}}$ and $\beta$ are parameters of the algorithm.
Successive splittings are evaluated until one passes the grooming condition, which tags the event, or no splittings remain (untagged event).
This analysis utilizes $z_{\text{cut}}$ = 0.2 and $\beta$ = 0 to preferentially select more symmetric splittings.

For dynamical grooming (``DyG'')~\cite{Mehtar-Tani:2019rrk}, the hardness measure is $\kappa \propto z(1-z)p_{\text{T,split}}(\Delta R/R)^{a}$, where $a$ is a tunable hardness parameter and $p_{\text{T,split}}$ is the transverse momentum of the parent splitting.
There is no explicit grooming condition, in contrast to SD; rather, the splitting with the largest value of hardness measure is always selected.
This analysis utilizes $a$ = 1.0 to preferentially select high-\kT{} splittings which may not be selected by SD~\cite{Caucal:2021cfb}.

Single hard scatterings may generate hard parton shower splittings.
In this analysis, jet grooming is used to search for such hard splittings using the distribution of the relative transverse momentum of each splitting,

\begin{equation}
    k_{\text{T}} = p_{\text{T,subleading}} \sin{\Delta R},
\end{equation}

where $p_{\text{T,subleading}}$ is the transverse momentum of the softer subjet, $\Delta R = \sqrt{\Delta y^{2} + \Delta \varphi^{2}}$, and $\Delta \varphi$ and $\Delta y$ are the subjet azimuthal angle and rapidity difference.
The splitting phase space is characterized by $\kT{}$ and $\deltaR$, which are used to construct the Lund Plane~\cite{Andrews:2018jcm}, and can be related by the momentum sharing fraction. The value of the splitting selected by the grooming algorithm is denoted $\kTg{}$.

The complex background in \PbPb{} collisions generates false jet shower branches which may mask physical jet splittings and the physical leading and subleading subjets, with the subleading subjet more susceptible to contamination~\cite{Mulligan:2020tim,ALargeIonColliderExperiment:2021mqf}.
The probability to falsely tag a background splitting is determined using embedding~\cite{Mulligan:2020tim}.
Measurements with small $R$ in non-central \PbPb{} collisions mitigate such effects.

The DyG algorithm is especially susceptible to selecting incorrectly identified subjets since it considers all iterative splittings, without imposing a hard cutoff in the grooming condition.
This effect is mitigated in the \PbPb{} analysis by requiring $\kTgMeas{} >$ 1 \GeVc{} in semicentral and $\kTgMeas{} >$ 1.5 \GeVc{} in central collisions, where $\kTgMeas{}$ is the background-subtracted but otherwise uncorrected \kTg{} distribution. This requirement trades more limited kinematic reach for increased purity; for details see Appendix~\ref{sec:subjetPurity}.

This minimum \kTgMeas{} requirement and the SD grooming condition similarly reduce soft contamination and improve tagging purity.
For central \PbPb{} collisions,
\sdZA{0.2} and \dygAA{1.0} with \kTgMeas{} have a subleading subjet true tagging rate of $\sim 80-90$\%.

The $\kTg{}$ distributions are corrected for background fluctuations and detector effects, including track $\pT{}$ resolution and tracking inefficiency.
A response matrix is constructed in \PbPb{} (\pp{}) collisions by correlating a probe jet at particle level and reconstructed level (detector level), relating the $\pTJet{}$ and $\kTg{}$ distributions.
The response matrix is used to perform 2D iterative Bayesian unfolding~\cite{DAgostini:1994fjx, DAgostini:2010hil} as implemented in RooUnfold~\cite{Adye:2011gm}.
The optimal iteration for unfolding regularization is chosen such that variation in the unfolded distribution for successive iterations is less than 1\%.
The iteration number for regularization, which is optimized for each grooming method and collision system, ranges from 6 to 11.

For \PbPb{} collisions, systematic uncertainty in the $\kTg{}$ distribution arises from the unfolding procedure, tracking efficiency, generator dependence of the response matrix, subtraction of the underlying event, and non-closure of the unfolding procedure due to mis-tagged subjets.
For \pp{} collisions, systematic uncertainty arises from unfolding, tracking efficiency, and generator dependence.

Systematic uncertainties are quantified as described in Appendix~\ref{sec:systematicUncertaintiesFull}.
The systematic uncertainty due to track selection and ITS--TPC track-matching efficiency uncertainty is determined by randomly rejecting tracks with a given probability, which is $\pT{}$-dependent in \PbPb{} collisions.
Under- and over-subtraction of background is assessed by varying the constituent subtraction distance parameter $R_{\text{max}}$.
Systematic uncertainties in the unfolding procedure are evaluated by varying the regularization parameter, the minimum jet $\pTJetCh{}$ range, the $\kTg{}$ binning, and the prior.
Generator dependence of the corrections is evaluated by unfolding with an alternative response matrix constructed using HERWIG 7~\cite{Bahr:2008pv, Bellm:2015jjp} in \pp{} collisions and JEWEL without recoils in \PbPb{} collisions.

A closure test which varies analysis components with respect to their nominal values is utilized to assess the full analysis chain.
Closure corresponds to consistency within statistical error of corrected spectra for the variant and nominal components.
Component variations are used to assess systematic uncertainties and analysis consistency, such as the sensitivity to the prior spectrum shape, as well as the residual background contribution via embedding and a parameterized thermal model.
Point-by-point excursions from the closure tests beyond $1\sigma$ statistical error contribute to the systematic uncertainty.
Closure tests to access the impact of residual background contamination exhibit non-closure of up to 12\% in a single bin, with largest non-closure in central \PbPb{} collisions.

The various contributions to the systematic uncertainty are largely uncorrelated among each other and are added in quadrature.
The cumulative systematic uncertainty has magnitude 8--32\% for central \PbPb{} and 2--7\% for \pp{} collisions.
The dominant uncertainties arise from the generator dependence of the corrections, subtraction, and non-closure for \PbPb{} collisions, and the generator dependence and tracking for \pp{} collisions.

\begin{figure}[ht]
    \centering
    \includegraphics[width=0.7\textwidth]{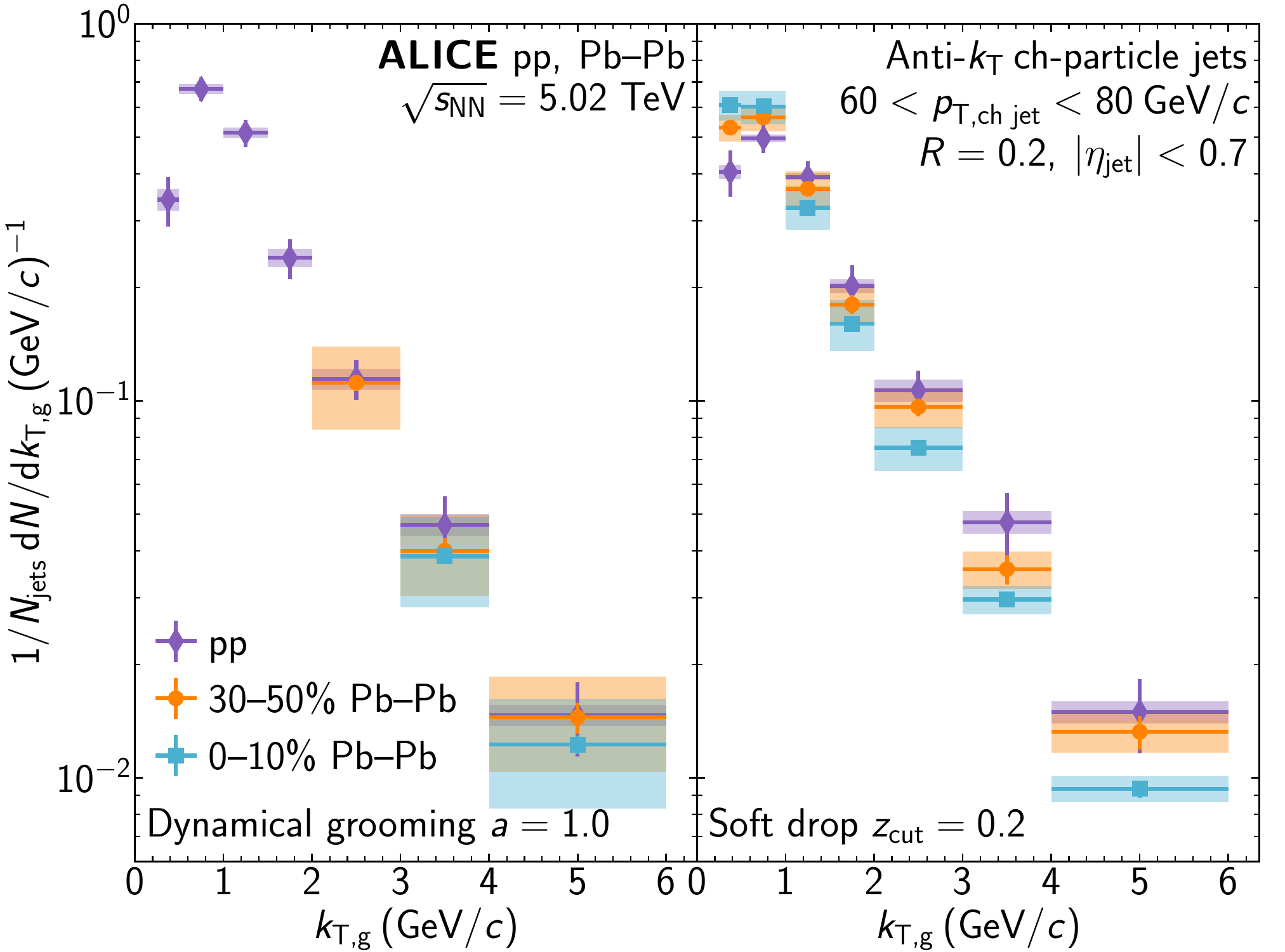}
    \caption{Unfolded $\kTg{}$ spectra for splittings selected by \dygA{1.0} (left) and \sdZ{0.2} (right) in \pp{} and semicentral and central \PbPb{} collisions. The jets are measured with $R=0.2$ in the $60 < \pTJetCh{} < 80$ \GeVc{} interval. Statistical uncertainties are represented by bars, while systematic uncertainties are shown in shaded boxes.}
    \label{fig:ktgSpectra}
\end{figure}

Figure~\ref{fig:ktgSpectra} shows the corrected $\kTg{}$ yield distributions measured in \pp{} and in central and semicentral \PbPb{} collisions, using DyG ($a=1.0$) (left) and SD ($z_{\text{cut}}=0.2$) (right).
Each distribution is normalized by the number of jets found in the $60 < \pTJetCh{} < 80$ \GeVc{} interval.
This is the first measurement of heavy-ion collision data to employ DyG.
All distributions have a similar shape, with a steeply decreasing slope for $\kTg{} > 1$ \GeVc{}.
The results from \pp{} collisions show a maximum at $\kTg{} \approx$ 0.5--1 \GeVc{}.
The \dygAA{1.0} spectra in \PbPb{} collisions are restricted to $\kTg{} > 2$~\GeVc{} ($\kTg{} > 3$~\GeVc{}) in semicentral (central) collisions due to the minimum $\kTg{}$ requirement discussed above.
The SD ($z_{\text{cut}}=0.2$) spectra from \pp{} collisions exhibit larger yield at high $\kTg{}$ than those from \PbPb\ collisions,
while for \dygAA{1.0}, the \pp{} and \PbPb{} spectra are consistent within uncertainties for $\kTg{} > 2$~\GeVc{} where the \PbPb{} results are available.
Comparing at fixed centrality and collision system, the \dygAA{1.0} and \sdZA{0.2} distributions are consistent over the measured $\kTg{}$ range within the systematic error and statistical uncertainty.

\begin{figure}[ht]
    \centering
    \includegraphics[width=0.65\textwidth]{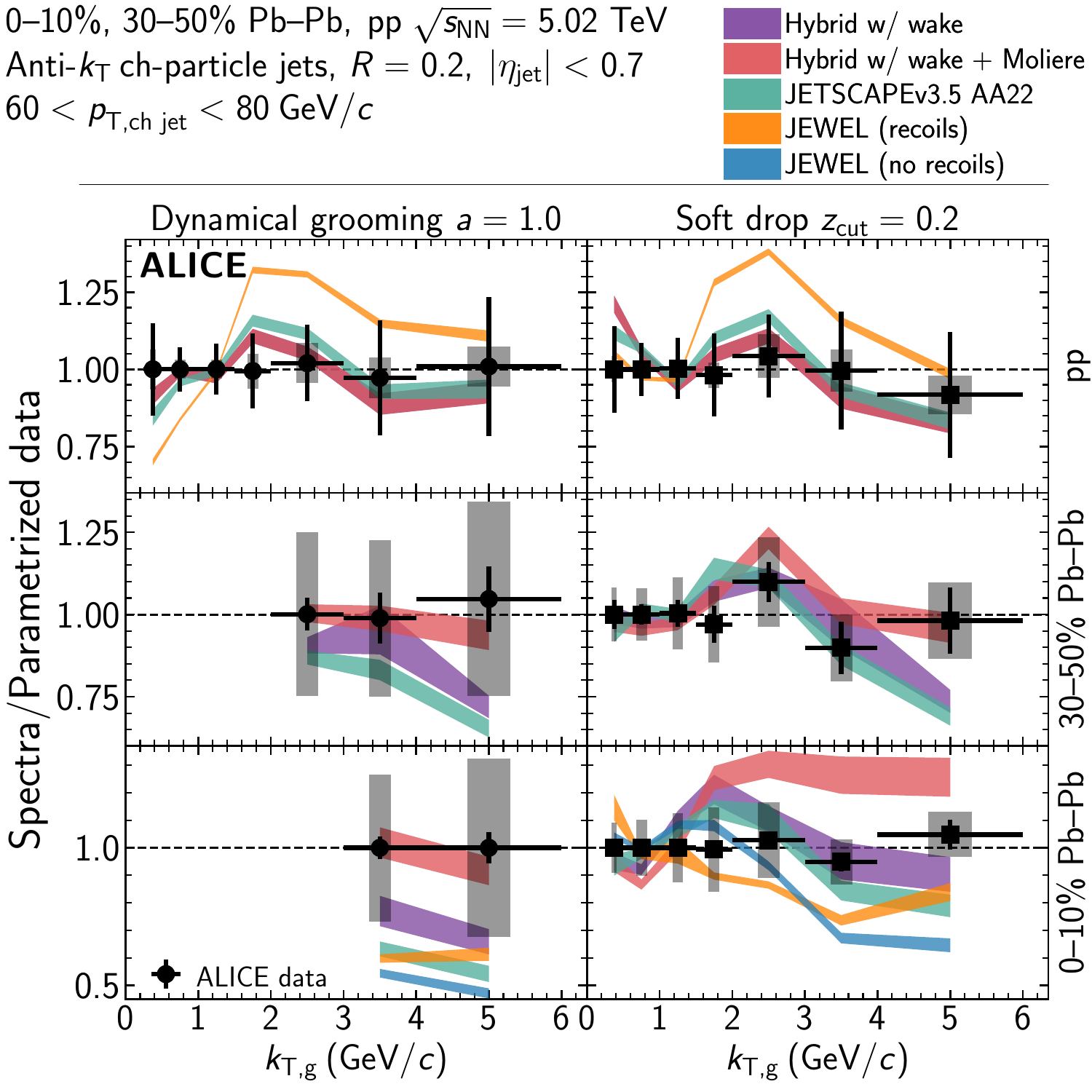}
    \caption{Ratio of $\kTg{}$ distributions for data and model calculations to a smooth parametrization of the data, selected by \dygA{1.0} (left) and \sdZ{0.2} (right). Data points are shown in black, with statistical (error bars) and systematic (grey boxes) uncertainties. The bands for the model calculations show statistical uncertainties.}
    \label{fig:ktgSpectraWithModelComparisons}
\end{figure}

Figure~\ref{fig:ktgSpectraWithModelComparisons} shows the same distributions as Fig.~\ref{fig:ktgSpectra}, in this case as the ratio with a smooth function that parametrizes the data (details in Appendix~\ref{sec:spectraParametrization}).
This representation removes the main functional dependence on $\kTg{}$ so that detailed variation of the distributions can be seen.
The function parameters are different for each collision system, centrality class, and grooming method.
Model calculations are also shown, and are divided by the same data parametrization.
In the upper panels, the central values of the JETSCAPE and Hybrid Model calculations tend to underpredict the measured \pp{} distribution at high $\kT{}$ for both \dygAA{1.0} and \sdZA{0.2}, though are consistent with the data within the uncertainties.
At low $\kTg{}$ ($< 1$ \GeVc{}), the JETSCAPE and Hybrid calculations with \sdZA{0.2} grooming hint towards slightly overpredicting the data, while JEWEL underpredicts the \dygAA{1.0} data.
JEWEL overestimates the data at intermediate \kTg{} for both grooming methods.
For semicentral \PbPb{} collisions, displayed in the middle panels, both the \dygAA{1.0} and \sdZA{0.2} measurements are described well by all models for $\kTg{} < 4$ \GeVc{}.
However, at high $\kTg{}$, the Hybrid Model with \moliere{} scattering (\hybridWithMoliere{}) agrees better with the \sdZA{0.2} data, while there is a hint that the other models underpredict the measured distributions.
The bottom panels show tension of the models and data for \sdZA{0.2} in central \PbPb{ collisions}.
At high $\kTg{} > 3$ \GeVc{}, the Hybrid Model without \moliere{} scattering (\hybridWithoutMoliere{}) best describes the data, while \hybridWithMoliere{} overpredicts, JETSCAPE slightly underpredicts, and JEWEL with and without recoils underpredict the measured values.
For \dygAA{1.0}, \hybridWithMoliere{} and \hybridWithoutMoliere{} best describe the data, while JETSCAPE and JEWEL underpredict the ratio.
The large systematic uncertainties of the \dygAA{1.0} spectra are due to a strong dependence of the grooming algorithm on the generator used to derive corrections of the spectra, limiting the model discrimination power.
This indicates that the additional phase space probed by the algorithm does not provide further resolving power.

\begin{figure}[ht]
    \centering
    \includegraphics[width=0.65\textwidth]{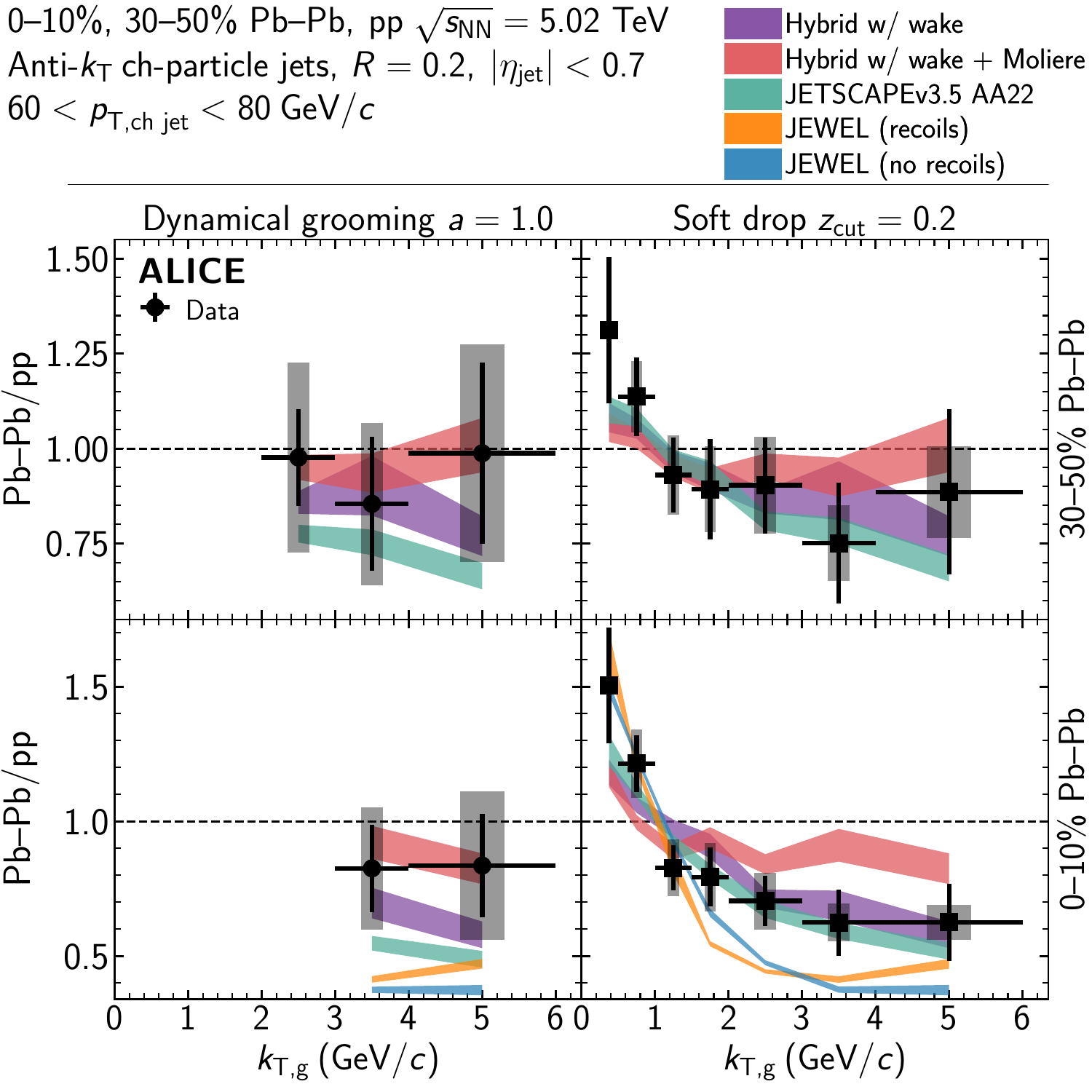}
    \caption{Ratio of the $\kTg{}$ distributions measured in \PbPb{} and \pp{} collisions for \dygA{1.0} (left) and \sdZA{0.2} (right), with comparison to model calculations. Statistical uncertainties are shown as error bars and systematic uncertainties are shown as grey shaded boxes for the measured ratios. The width of the colored bands corresponds to the statistical uncertainties in the model calculations.}
    \label{fig:ktgRatiosWithModelComparisons}
\end{figure}

In-medium \moliere{} scattering is expected to generate yield enhancement at large $\kTg{}$~\cite{DEramo:2012uzl,DEramo:2018eoy}, as seen in the Hybrid model variants in Fig~\ref{fig:ktgSpectraWithModelComparisons}.
To search for this effect, Fig.~\ref{fig:ktgRatiosWithModelComparisons} shows the ratios of $\kTg{}$ distributions for semicentral and central \PbPb{} collisions to that for \pp{} collisions.
The \sdZA{0.2} distributions exhibit medium-induced yield enhancement for $\kTg{} < 1$ \GeVc{} and yield suppression at higher \kTg{}, with stronger medium-induced effects in central than in semicentral \PbPb{} collisions.
The \dygAA{1.0} distributions hint at yield suppression at large $\kTg{}$ for central \PbPb{} collisions but not for semicentral \PbPb{} collisions, though the ratios in both cases are consistent with unity within the systematic uncertainties.
Since yield suppression is observed in the data in the high-$\kTg{}$ region, this measurement does not show direct evidence of in-medium \moliere{} scattering.
The \sdZA{0.2} analysis for central \PbPb{} collisions~(Fig.~\ref{fig:ktgRatiosWithModelComparisons}, lower right) exhibits the most significant medium-induced modifications in this analysis, with experimental uncertainties which provide the best discrimination of model predictions. We therefore focus primarily on these data for detailed model comparisons and physics conclusions. 

For \sdZA{0.2} the central \PbPb{}/pp{} yield ratio is smaller than unity (where unity corresponds to the null hypothesis) by $2.7\sigma$ for $3<\kTg{}<4$ \GeVc{} and by $2.4\sigma$ for $4<\kTg{}<6$ \GeVc{} based on a quadrature sum of the statistical error and systematic uncertainty.
This medium-induced narrowing of jet structure is qualitatively consistent with effects found in other inclusive jet substructure measurements~\cite{ALargeIonColliderExperiment:2021mqf,ATLAS:2022vii,ALICE:2023dwg}, including the trend given by the combination of the groomed jet radius and momentum sharing.
Such narrowing is commonly interpreted as selection bias which modifies the relative fraction of quark- or gluon-originating jets, due to the combined effects of a falling \pT{} spectrum and larger energy loss in the medium for wider jets~\cite{Casalderrey-Solana:2016jvj,Ringer:2019rfk,Du:2020pmp,Kang:2023ycg}. 
The smaller modification of the \kTg{} distribution observed in semicentral \PbPb{} collisions is consistent with this picture, due to the correspondingly smaller jet energy loss in that population.
The single-hard scattering signature may also be suppressed by subsequent multiple soft interactions with the medium, which smear the angular correlations of jet constituents, such that the \moliere{} scattering is not reconstructed~\cite{Mehtar-Tani:2022zwf}.
Medium modification of coherence length~\cite{Hulcher:2017cpt,Casalderrey-Solana:2019ubu} and of emission formation time~\cite{Apolinario:2024hsm} may also explain the measured narrowing.

Figure~\ref{fig:ktgRatiosWithModelComparisons} also shows model calculations. 
None of the models describe all measured distributions in Figs.~\ref{fig:ktgSpectraWithModelComparisons} and~\ref{fig:ktgRatiosWithModelComparisons}, with different levels of agreement for the two grooming algorithms. 
The central \PbPb{} data (lower right) are well-described within uncertainties by JETSCAPE (with \moliere{} scattering) and \hybridWithoutMoliere{}, but not by \hybridWithMoliere{} or either variant of JEWEL.
From this comparison it is evident that the different theoretical approaches of these models to jet quenching and the bulk dynamics of the QGP generate markedly different predictions for the relative effect of \moliere{} scattering, which these and other complementary measurements~\cite{ALICE:2023qve} can discriminate. Further exploration of the key model features underlying these differences requires a dedicated global analysis of the world’s jet quenching data (e.g.~\cite{JETSCAPE:2024cqe}).

This Letter reports the first measurement of the distribution of the hardest relative transverse jet splitting, $\kTg{}$, in \pp{} and \PbPb{} collisions, to search for evidence of quasi-particle scattering in the quark--gluon plasma.
The hardest splitting is identified using the soft drop and dynamical grooming algorithms, the latter for the first time in heavy-ion collisions.
The high-$\kTg{}$ yield in central \PbPb{} collisions is observed to be suppressed relative to that in \pp{} collisions, consistent with in-medium narrowing effects seen in other jet substructure observables.
No direct evidence is observed for in-medium quasi-particle scattering within the measured phase space. Model calculations can reproduce the measurements within uncertainties, but differ in the necessity to include quasi-particle scattering effects in order to do so. In these models, the \moliere{} scattering signal may be masked by yield suppression due to jet energy loss, with their relative effects depending upon the theoretical picture underlying the modeling of jet quenching, bulk dynamics, and the vacuum parton shower. These data therefore provide new constraints on theoretical modeling of the microscopic structure and dynamics of the quark--gluon plasma.

%%%%%%%%%%%%%%%%%%%%%%%%%%%%%%%%
% end main text
%%%%%%%%%%%%%%%%%%%%%%%%%%%%%%%%

%%%%% acknowledgements - handled by EB chairs
\newenvironment{acknowledgement}{\relax}{\relax}
\begin{acknowledgement}
\section*{Acknowledgements}
% add specific acknowledgements here
We gratefully acknowledge Zach Hulcher and Daniel Pablos, and Yasuki Tachibana and the JETSCAPE collaboration, for providing calculations.
% ...but please don't remove the line below: funding agencies
% will be acknowledged with a custom tex file handled by EB chairs after Collab Round 2
% Version: 2024-08-29

The ALICE Collaboration would like to thank all its engineers and technicians for their invaluable contributions to the construction of the experiment and the CERN accelerator teams for the outstanding performance of the LHC complex.
The ALICE Collaboration gratefully acknowledges the resources and support provided by all Grid centres and the Worldwide LHC Computing Grid (WLCG) collaboration.
The ALICE Collaboration acknowledges the following funding agencies for their support in building and running the ALICE detector:
A. I. Alikhanyan National Science Laboratory (Yerevan Physics Institute) Foundation (ANSL), State Committee of Science and World Federation of Scientists (WFS), Armenia;
Austrian Academy of Sciences, Austrian Science Fund (FWF): [M 2467-N36] and Nationalstiftung f\"{u}r Forschung, Technologie und Entwicklung, Austria;
Ministry of Communications and High Technologies, National Nuclear Research Center, Azerbaijan;
Conselho Nacional de Desenvolvimento Cient\'{\i}fico e Tecnol\'{o}gico (CNPq), Financiadora de Estudos e Projetos (Finep), Funda\c{c}\~{a}o de Amparo \`{a} Pesquisa do Estado de S\~{a}o Paulo (FAPESP) and Universidade Federal do Rio Grande do Sul (UFRGS), Brazil;
Bulgarian Ministry of Education and Science, within the National Roadmap for Research Infrastructures 2020-2027 (object CERN), Bulgaria;
Ministry of Education of China (MOEC) , Ministry of Science \& Technology of China (MSTC) and National Natural Science Foundation of China (NSFC), China;
Ministry of Science and Education and Croatian Science Foundation, Croatia;
Centro de Aplicaciones Tecnol\'{o}gicas y Desarrollo Nuclear (CEADEN), Cubaenerg\'{\i}a, Cuba;
Ministry of Education, Youth and Sports of the Czech Republic, Czech Republic;
The Danish Council for Independent Research | Natural Sciences, the VILLUM FONDEN and Danish National Research Foundation (DNRF), Denmark;
Helsinki Institute of Physics (HIP), Finland;
Commissariat \`{a} l'Energie Atomique (CEA) and Institut National de Physique Nucl\'{e}aire et de Physique des Particules (IN2P3) and Centre National de la Recherche Scientifique (CNRS), France;
Bundesministerium f\"{u}r Bildung und Forschung (BMBF) and GSI Helmholtzzentrum f\"{u}r Schwerionenforschung GmbH, Germany;
General Secretariat for Research and Technology, Ministry of Education, Research and Religions, Greece;
National Research, Development and Innovation Office, Hungary;
Department of Atomic Energy Government of India (DAE), Department of Science and Technology, Government of India (DST), University Grants Commission, Government of India (UGC) and Council of Scientific and Industrial Research (CSIR), India;
National Research and Innovation Agency - BRIN, Indonesia;
Istituto Nazionale di Fisica Nucleare (INFN), Italy;
Japanese Ministry of Education, Culture, Sports, Science and Technology (MEXT) and Japan Society for the Promotion of Science (JSPS) KAKENHI, Japan;
Consejo Nacional de Ciencia (CONACYT) y Tecnolog\'{i}a, through Fondo de Cooperaci\'{o}n Internacional en Ciencia y Tecnolog\'{i}a (FONCICYT) and Direcci\'{o}n General de Asuntos del Personal Academico (DGAPA), Mexico;
Nederlandse Organisatie voor Wetenschappelijk Onderzoek (NWO), Netherlands;
The Research Council of Norway, Norway;
Pontificia Universidad Cat\'{o}lica del Per\'{u}, Peru;
Ministry of Science and Higher Education, National Science Centre and WUT ID-UB, Poland;
Korea Institute of Science and Technology Information and National Research Foundation of Korea (NRF), Republic of Korea;
Ministry of Education and Scientific Research, Institute of Atomic Physics, Ministry of Research and Innovation and Institute of Atomic Physics and Universitatea Nationala de Stiinta si Tehnologie Politehnica Bucuresti, Romania;
Ministry of Education, Science, Research and Sport of the Slovak Republic, Slovakia;
National Research Foundation of South Africa, South Africa;
Swedish Research Council (VR) and Knut \& Alice Wallenberg Foundation (KAW), Sweden;
European Organization for Nuclear Research, Switzerland;
Suranaree University of Technology (SUT), National Science and Technology Development Agency (NSTDA) and National Science, Research and Innovation Fund (NSRF via PMU-B B05F650021), Thailand;
Turkish Energy, Nuclear and Mineral Research Agency (TENMAK), Turkey;
National Academy of  Sciences of Ukraine, Ukraine;
Science and Technology Facilities Council (STFC), United Kingdom;
National Science Foundation of the United States of America (NSF) and United States Department of Energy, Office of Nuclear Physics (DOE NP), United States of America.
In addition, individual groups or members have received support from:
Czech Science Foundation (grant no. 23-07499S), Czech Republic;
FORTE project, reg.\ no.\ CZ.02.01.01/00/22\_008/0004632, Czech Republic, co-funded by the European Union, Czech Republic;
European Research Council (grant no. 950692), European Union;
ICSC - Centro Nazionale di Ricerca in High Performance Computing, Big Data and Quantum Computing, European Union - NextGenerationEU;
Academy of Finland (Center of Excellence in Quark Matter) (grant nos. 346327, 346328), Finland.

\end{acknowledgement}

%%%%%%%% Bibliography
\bibliographystyle{utphys}   % Remember we use title in the biblio
\bibliography{bibliography}
%\input {bibliography.tex}

%%%%%%%%%%%%%%%%%%%%%%%%%%%%%%%%
% Appendices: yours (if any) + authorlist
%%%%%%%%%%%%%%%%%%%%%%%%%%%%%%%%
\newpage
\appendix

% Appendixes here
\section*{End Matter}
\section{Subjet purity}\label{sec:subjetPurity}

The large underlying event present in heavy-ion collisions contaminates the jet shower, leading to potential misidentification of physical jet splittings and the contributing subjets~\cite{Mulligan:2020tim,ALargeIonColliderExperiment:2021mqf}.
For example, a true leading subjet may be paired with an uncorrelated jet constituent which is coincidentally close in angular phase space, leading to no hard splittings being identified in the declustering tree.
Unlike algorithms with a hard cutoff in the grooming condition, DyG requires that a splitting be identified for each measured jet, even in the case that a jet is solely composed of soft splittings.
The heavy background contamination of such soft splittings causes unfolding instabilities.
These instabilities may be addressed via modifications to existing analysis procedures, which are explored below.

\begin{figure}[ht]
    \centering
    \includegraphics[width=0.9\textwidth]{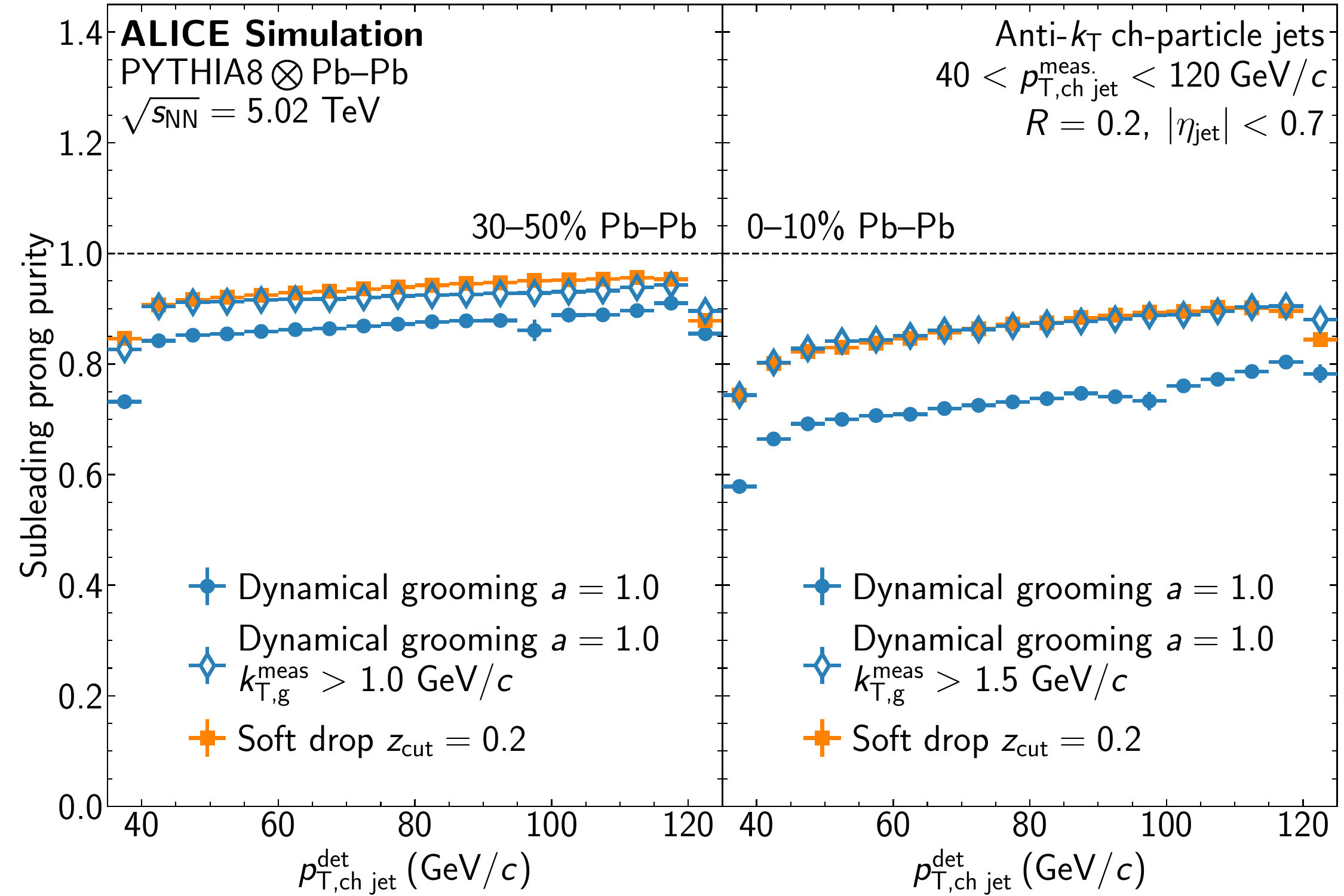}
    \caption{Subleading subjet purity for \sdZ{0.2} and \dygA{1.0} applied to jets found in semicentral (left) and central (right) \PbPb{} collisions. Closed markers correspond to the standard grooming algorithms, while open markers represent \dygA{1.0} with a minimum $\kTgMeas{}$ requirement. Statistical uncertainties are represented by bars.}
    \label{fig:subleadingSubjetPurity}
\end{figure}

To characterize the impact of background contamination and study mitigation techniques, PYTHIA 8 probe jets generated with the Monash 2013 tune are embedded into \PbPb{} data, following a similar evaluation procedure as described in Ref.~\cite{Mulligan:2020tim}.
The grooming algorithms are applied to the probe jets at both the detector and reconstructed (i.e. embedded) levels.
Performance is characterized by the purity, which is the fraction of subjets which are correctly identified in reconstructed-level jets, normalized by the number of groomed subjets found in detector-level jets.
Identified splittings and subjets are matched based on subjet constituents, with the reconstructed-level subjet required to carry at least 50\% of the detector-level subjet $\pT{}$.
The leading subjet prong is correctly identified in more than 95\% of all splittings, and is not shown here.
The subleading subjet prong purity as a function of the detector-level probe $\pTJetCh{}$ is shown in Fig.~\ref{fig:subleadingSubjetPurity} for semicentral (left) and central (right) \PbPb{} collisions.
Detector-level \pTJetCh{} is utilized to assess the impact of background on the purity (as compared to detector effects). The $40 < p_{\text{T,ch jet}}^{\text{det}} < 120$ \GeVc{} interval contains nearly all of the particle-level jets which contribute to the unfolded measurement.
Soft drop \zcut{0.2} correctly identifies more than 80\% (90\%) of subleading prongs in central (semicentral) \PbPb{} collisions.
However, the purity drops to 65\% (85\%) for \dygA{1.0}, which leads to unfolding instabilities.

The subleading subjet purity can be improved by excluding the most background contaminated region of the $\kTg{}$ spectra by requiring that the background-subtracted but otherwise uncorrected value, $\kTgMeas{}$, is larger than a minimum value.
Figure~\ref{fig:subleadingSubjetPurity}, open markers, demonstrates this effect, with the subleading subjet purity of \dygA{1.0} with a minimum $\kTgMeas{}$ requirement increasing to near agreement with \sdZ{0.2}.
This requirement causes a corresponding reduction in the kinematic efficiency, which describes the efficiency of finding particle-level jets given a fixed detector-level kinematic interval~\cite{ALICE:2013dpt}, at low $\kTg{} \sim$ 1--2 \GeVc{}.
Reduced kinematic efficiency restricts the reported kinematic range to larger $\kTg{}$ values where the kinematic efficiency is relatively higher, and increases sensitivity to the generator used to correct for this reduction.
This analysis requires a kinematic efficiency greater than $\sim 70\%$.
Optimizing the increase in purity against the reduction in kinematic efficiency (and therefore the reported kinematic range), $\kTgMeas{} > 1.5$  ($1.0$) \GeVc{} in central (semicentral) \PbPb{} collisions was selected for this measurement.
The performance of JEWEL without recoils is consistent with the values shown in Fig.~\ref{fig:subleadingSubjetPurity}, indicating limited model sensitivity.
Imposing this $\kTgMeas$ condition for \sdZ{0.2} does not improve the subleading subjet purity because the grooming condition already removes the background contamination.

\section{Systematic uncertainties}\label{sec:systematicUncertaintiesFull}

The systematic uncertainties in this measurement are due to the tracking efficiency ("trk. eff."), the subtraction of the underlying background ("bkgd. sub"), 
the unfolding procedure, the generator dependence of the unfolding procedure, and the residual non-closure of the unfolding procedure.
Only the tracking, generator dependence, and unfolding apply to the measurement in \pp{} collisions.
Sources of uncertainty are evaluated following similar procedures and for similar physical considerations as described in Ref.~\cite{ALargeIonColliderExperiment:2021mqf}.
Signed uncertainties are symmetrized.
The total uncertainty is summed in quadrature over the sources mentioned above, assuming no correlation between the individual uncertainties.

Track selections are considered along with the uncertainty on the ITS--TPC track-matching efficiency to estimate the systematic uncertainty on the observable due to the tracking performance.
In \pp{} collisions, this amounts to a \pT{}-independent uncertainty of $\sim 3\%$, while in \PbPb{} collisions, a \pT{}-dependent parametrization peaks at $\sim 6\%$ for $\pT{} = 1$ \GeVc{} and decreases to $\sim 3\%$ for $\pT{} \geq 15$ \GeVc{}.
This performance dependence is assessed by randomly rejecting tracks according to the magnitude of the uncertainties.

Imperfect background subtraction may bias the $\pTJetCh{}$ and $\kT{}$ dependence of the measured $\kTg{}$ distributions.
The analysis was carried out with two values of the constituent subtraction distance parameter $R_{\text{max}}$, 0.05 and 0.5, to characterize under- and over-subtraction of background contributions.
The maximum bin-by-bin variation is assigned as a systematic uncertainty.

Sensitivity to the unfolding procedure is evaluated via variations of the unfolding parameters, including:
varying the regularization parameter of the Bayesian unfolding by $\pm 1$ unit;
increasing and decreasing the minimum smeared jet $\pTJetCh{}$ range by $\pm 5$ \GeVc{} in \PbPb{} collisions ($\pm 3$ \GeVc{} in \pp{} collisions);
modifying the $\kTg{}$ bin edges by up to $\pm 5\%$;
and weighing the prior distribution to resemble the uncorrected raw distributions.
These variations are correlated in their characterization of the unfolding performance and sensitivity.
Consequently, the variations are combined into a single unfolding uncertainty, calculated as the standard deviation of the individual values.

The generator utilized to construct the response matrix used for the unfolding can impact the measured distributions.
This is evaluated by constructing alternative response matrices using HERWIG 7~\cite{Bahr:2008pv, Bellm:2015jjp} in \pp{} collisions, and JEWEL 2.2.0~\cite{Zapp:2012ak,Zapp:2013vla} without recoils in \PbPb{} collisions.
A fast simulation is employed as described in Ref.~\cite{ALargeIonColliderExperiment:2021mqf}, and the uncertainty is determined as the maximum bin-by-bin variation.
The systematic uncertainty derived in central collisions is also applied in semicentral collisions.

To account for the effects of residual subjet misidentifications, two methods are considered.
Closure tests are performed using two approaches: 1) unfolding a known input distribution which is weighed to resemble the measured data;
and 2) unfolding a known probe distribution which is embedded into a thermal background, as described in Ref.~\cite{ALargeIonColliderExperiment:2021mqf}.
Any remaining non-closure in these correction procedures is taken as a systematic uncertainty, with the value assigned as the maximum bin-by-bin difference of the two methods from the nominal value.

Relative systematic uncertainties of $\kTg{}$ distributions measured using \sdZ{0.2} and \dygA{1.0} are shown in
Table~\ref{tab:ppSystematicUncertainties} for \pp{} collisions and
Table~\ref{tab:PbPbSystematicUncertainties} for 0--10\% and 30--50\% \PbPb{} collisions.

% NOTE: Values are generated by pandas table.
%       The only manual edit is to add the top line of the table - i.e.:
%       \pp{} & \multicolumn{4}{c}{Relative uncertainty (\%)} \\
%       \midrule
\begin{table}[ht]
    \centering
    \caption{Relative systematic uncertainties (\%) for $\kTg{}$ distributions measured using \sdZ{0.2} and \dygA{1.0} in \pp{} collisions. The minimum and maximum values over the measured $\kTg{}$ range are reported. The combined uncertainty is the sum in quadrature of the individual uncertainties.}
    \begin{tabular}{lllll}
    \toprule
    \pp{} & \multicolumn{4}{c}{Relative uncertainty (\%)} \\
    \midrule
     & Trk. Eff. & Generator & Unfolding & Total \\
    \midrule
    DyG. $a$ = 1.0 & 1 -- 6\% & 2 -- 6\% & 1 -- 4\% & 3 -- 7\% \\
    SD $z_{\mathrm{cut}}$ = 0.2 & 0 -- 4\% & 1 -- 6\% & 1 -- 2\% & 2 -- 7\% \\
    \bottomrule
    \end{tabular}
    \label{tab:ppSystematicUncertainties}
\end{table}

% NOTE: Values are generated by pandas table.
%       The only manual edit is to add the top line of the table - i.e.:
%       \PbPb{} & \multicolumn{7}{c}{Relative uncertainty (\%)} \\
%       \midrule
\begin{table}[ht]
    \centering
    \caption{Relative systematic uncertainties (\%) for $\kTg{}$ distributions measured using \sdZ{0.2} and \dygA{1.0} in central and semicentral \PbPb{} collisions. The minimum and maximum values over the measured $\kTg{}$ range are reported. The combined uncertainty is the sum in quadrature of the individual uncertainties.}
    \begin{tabular}{llllllll}
    \toprule
    \PbPb{} & \multicolumn{7}{c}{Relative uncertainty (\%)} \\
    \midrule
     &  & Trk. Eff. & Bkgd. Sub. & Unfolding & Generator & Non-closure & Total \\
    \midrule
    DyG. $a = 1.0$ & 0--10\% & 4 -- 7\% & 8 -- 14\% & 1 -- 2\% & 22 -- 26\% & 12 -- 12\% & 27 -- 32\% \\
    SD $z_{\mathrm{cut}} = 0.2$ & 0--10\% & 0 -- 7\% & 1 -- 7\% & 0 -- 2\% & 2 -- 11\% & 2 -- 6\% & 8 -- 15\% \\
    \rule{0pt}{3ex}DyG. $a = 1.0$ & 30--50\% & 1 -- 6\% & 4 -- 9\% & 1 -- 2\% & 22 -- 26\% & 2 -- 5\% & 24 -- 28\% \\
    SD $z_{\mathrm{cut}} = 0.2$ & 30--50\% & 0 -- 5\% & 1 -- 9\% & 0 -- 2\% & 2 -- 11\% & 1 -- 5\% & 8 -- 12\% \\
    \bottomrule
    \end{tabular}
    \label{tab:PbPbSystematicUncertainties}
\end{table}

\section{Parametrization of spectra}\label{sec:spectraParametrization}

A smoothed parametrization was employed in order to factorize the overall features of the spectra while separating out the fluctuations of the data and model calculations.
The smoothed functions were separately fit to the data for each grooming method in \pp{}, semicentral, and central \PbPb{} collisions.
The same extracted parameters are used in the parametrized function to calculate the ratios with the data and the model calculations.
The spectra are parametrized by

\begin{equation}
    \begin{aligned}
    f(\kT{}, k_{\text{T}}^{0}, s) = [q(\kT{}, A, B) + (E - q(k_{\text{T}}^{0}, A, B))] t(\kT{}, k_{\text{T}}^{0}, -s)\\
    + [p(\kT{}, C, D) + (E - p(k_{\text{T}}^{0}, C, D))] t(\kT{}, k_{\text{T}}^{0}, s),
    \end{aligned}
    \label{eq:parametrization}
\end{equation}

with each term defined as,

\begin{equation}
    \begin{gathered}
    q(\kT{}; A, B) = A (\kT{} - B)^{2},\\
    p(\kT{}; C, D) = (C \kT{})^{-D},\\
    t(\kT{}; k_{\text{T}}^{0}, s) = (1 + \tanh((\kT{} - k_{\text{T}}^{0}) / s)) / 2,
    \end{gathered}
    \label{eq:parametrizationTerms}
\end{equation}

where $A$, $B$, $C$, $D$, and $E$ are fit parameters, $s = 0.25$ \GeVc{} in \pp{} collisions and $s = 0.1$ \GeVc{} in \PbPb{} collisions for both grooming methods, and $k_{\text{T}}^{0} = 1.25$ \GeVc{}.
For \dygAA{1.0} in semicentral and central \PbPb{} collisions where there are insufficient points to constrain $f(\kT{}, k_{\text{T}}^{0}, s)$, the data are fit to $p(\kT{}; C, D)$ alone.
The particular functional forms are arbitrary; they were selected to provide a smooth approximate description of the bulk features of the spectra, with a peak at low $\kTg{}$ via a quadratic term, transitioning to a power law at high $\kTg{}$.
The $1 + \tanh(...)$ element smoothly transitions between the two terms, while the normalization ensures that each term only contributes in its region of applicability.

%%%%% Authorlist - please do not touch: handled by EB chairs
\newpage
\section{The ALICE Collaboration}
\label{app:collab}
% ALICE Collaboration author list for 2024-08-29
\begin{flushleft} 
\small

S.~Acharya\,\orcidlink{0000-0002-9213-5329}\,$^{\rm 127}$, 
A.~Agarwal$^{\rm 135}$, 
G.~Aglieri Rinella\,\orcidlink{0000-0002-9611-3696}\,$^{\rm 32}$, 
L.~Aglietta\,\orcidlink{0009-0003-0763-6802}\,$^{\rm 24}$, 
M.~Agnello\,\orcidlink{0000-0002-0760-5075}\,$^{\rm 29}$, 
N.~Agrawal\,\orcidlink{0000-0003-0348-9836}\,$^{\rm 25}$, 
Z.~Ahammed\,\orcidlink{0000-0001-5241-7412}\,$^{\rm 135}$, 
S.~Ahmad\,\orcidlink{0000-0003-0497-5705}\,$^{\rm 15}$, 
S.U.~Ahn\,\orcidlink{0000-0001-8847-489X}\,$^{\rm 71}$, 
I.~Ahuja\,\orcidlink{0000-0002-4417-1392}\,$^{\rm 36}$
% ...

\section*{Affiliation Notes}

$^{\rm I}$ Deceased\\
$^{\rm II}$ Also at: Max-Planck-Institut fur Physik, Munich, Germany\\
$^{\rm III}$ Also at: Italian National Agency for New Technologies, Energy and Sustainable Economic Development (ENEA), Bologna, Italy\\
$^{\rm IV}$ Also at: Dipartimento DET del Politecnico di Torino, Turin, Italy\\
$^{\rm V}$ Also at: Department of Applied Physics, Aligarh Muslim University, Aligarh, India\\
$^{\rm VI}$ Also at: Institute of Theoretical Physics, University of Wroclaw, Poland\\
$^{\rm VII}$ Also at: Facultad de Ciencias, Universidad Nacional Autónoma de México, Mexico City, Mexico\\
$^{\rm VIII}$ Also at: An institution covered by a cooperation agreement with CERN\\

\section*{Collaboration Institutes}

$^{1}$ A.I. Alikhanyan National Science Laboratory (Yerevan Physics Institute) Foundation, Yerevan, Armenia\\
$^{2}$ AGH University of Krakow, Cracow, Poland\\
$^{3}$ Bogolyubov Institute for Theoretical Physics, National Academy of Sciences of Ukraine, Kiev, Ukraine\\
$^{4}$ Bose Institute, Department of Physics  and Centre for Astroparticle Physics and Space Science (CAPSS), Kolkata, India\\
$^{5}$ California Polytechnic State University, San Luis Obispo, California, United States\\
$^{6}$ Central China Normal University, Wuhan, China\\
$^{7}$ Centro de Aplicaciones Tecnol\'{o}gicas y Desarrollo Nuclear (CEADEN), Havana, Cuba\\
$^{8}$ Centro de Investigaci\'{o}n y de Estudios Avanzados (CINVESTAV), Mexico City and M\'{e}rida, Mexico\\
$^{9}$ Chicago State University, Chicago, Illinois, United States\\
$^{10}$ China Institute of Atomic Energy, Beijing, China\\
$^{11}$ China University of Geosciences, Wuhan, China\\
$^{12}$ Chungbuk National University, Cheongju, Republic of Korea\\
$^{13}$ Comenius University Bratislava, Faculty of Mathematics, Physics and Informatics, Bratislava, Slovak Republic\\
$^{14}$ Creighton University, Omaha, Nebraska, United States\\
$^{15}$ Department of Physics, Aligarh Muslim University, Aligarh, India\\
$^{16}$ Department of Physics, Pusan National University, Pusan, Republic of Korea\\
$^{17}$ Department of Physics, Sejong University, Seoul, Republic of Korea\\
$^{18}$ Department of Physics, University of California, Berkeley, California, United States\\
$^{19}$ Department of Physics, University of Oslo, Oslo, Norway\\
$^{20}$ Department of Physics and Technology, University of Bergen, Bergen, Norway\\
$^{21}$ Dipartimento di Fisica, Universit\`{a} di Pavia, Pavia, Italy\\
$^{22}$ Dipartimento di Fisica dell'Universit\`{a} and Sezione INFN, Cagliari, Italy\\
$^{23}$ Dipartimento di Fisica dell'Universit\`{a} and Sezione INFN, Trieste, Italy\\
$^{24}$ Dipartimento di Fisica dell'Universit\`{a} and Sezione INFN, Turin, Italy\\
$^{25}$ Dipartimento di Fisica e Astronomia dell'Universit\`{a} and Sezione INFN, Bologna, Italy\\
$^{26}$ Dipartimento di Fisica e Astronomia dell'Universit\`{a} and Sezione INFN, Catania, Italy\\
$^{27}$ Dipartimento di Fisica e Astronomia dell'Universit\`{a} and Sezione INFN, Padova, Italy\\
$^{28}$ Dipartimento di Fisica `E.R.~Caianiello' dell'Universit\`{a} and Gruppo Collegato INFN, Salerno, Italy\\
$^{29}$ Dipartimento DISAT del Politecnico and Sezione INFN, Turin, Italy\\
$^{30}$ Dipartimento di Scienze MIFT, Universit\`{a} di Messina, Messina, Italy\\
$^{31}$ Dipartimento Interateneo di Fisica `M.~Merlin' and Sezione INFN, Bari, Italy\\
$^{32}$ European Organization for Nuclear Research (CERN), Geneva, Switzerland\\
$^{33}$ Faculty of Electrical Engineering, Mechanical Engineering and Naval Architecture, University of Split, Split, Croatia\\
$^{34}$ Faculty of Nuclear Sciences and Physical Engineering, Czech Technical University in Prague, Prague, Czech Republic\\
$^{35}$ Faculty of Physics, Sofia University, Sofia, Bulgaria\\
$^{36}$ Faculty of Science, P.J.~\v{S}af\'{a}rik University, Ko\v{s}ice, Slovak Republic\\
$^{37}$ Faculty of Technology, Environmental and Social Sciences, Bergen, Norway\\
$^{38}$ Frankfurt Institute for Advanced Studies, Johann Wolfgang Goethe-Universit\"{a}t Frankfurt, Frankfurt, Germany\\
$^{39}$ Fudan University, Shanghai, China\\
$^{40}$ Gangneung-Wonju National University, Gangneung, Republic of Korea\\
$^{41}$ Gauhati University, Department of Physics, Guwahati, India\\
$^{42}$ Helmholtz-Institut f\"{u}r Strahlen- und Kernphysik, Rheinische Friedrich-Wilhelms-Universit\"{a}t Bonn, Bonn, Germany\\
$^{43}$ Helsinki Institute of Physics (HIP), Helsinki, Finland\\
$^{44}$ High Energy Physics Group,  Universidad Aut\'{o}noma de Puebla, Puebla, Mexico\\
$^{45}$ Horia Hulubei National Institute of Physics and Nuclear Engineering, Bucharest, Romania\\
$^{46}$ HUN-REN Wigner Research Centre for Physics, Budapest, Hungary\\
$^{47}$ Indian Institute of Technology Bombay (IIT), Mumbai, India\\
$^{48}$ Indian Institute of Technology Indore, Indore, India\\
$^{49}$ INFN, Laboratori Nazionali di Frascati, Frascati, Italy\\
$^{50}$ INFN, Sezione di Bari, Bari, Italy\\
$^{51}$ INFN, Sezione di Bologna, Bologna, Italy\\
$^{52}$ INFN, Sezione di Cagliari, Cagliari, Italy\\
$^{53}$ INFN, Sezione di Catania, Catania, Italy\\
$^{54}$ INFN, Sezione di Padova, Padova, Italy\\
$^{55}$ INFN, Sezione di Pavia, Pavia, Italy\\
$^{56}$ INFN, Sezione di Torino, Turin, Italy\\
$^{57}$ INFN, Sezione di Trieste, Trieste, Italy\\
$^{58}$ Inha University, Incheon, Republic of Korea\\
$^{59}$ Institute for Gravitational and Subatomic Physics (GRASP), Utrecht University/Nikhef, Utrecht, Netherlands\\
$^{60}$ Institute of Experimental Physics, Slovak Academy of Sciences, Ko\v{s}ice, Slovak Republic\\
$^{61}$ Institute of Physics, Homi Bhabha National Institute, Bhubaneswar, India\\
$^{62}$ Institute of Physics of the Czech Academy of Sciences, Prague, Czech Republic\\
$^{63}$ Institute of Space Science (ISS), Bucharest, Romania\\
$^{64}$ Institut f\"{u}r Kernphysik, Johann Wolfgang Goethe-Universit\"{a}t Frankfurt, Frankfurt, Germany\\
$^{65}$ Instituto de Ciencias Nucleares, Universidad Nacional Aut\'{o}noma de M\'{e}xico, Mexico City, Mexico\\
$^{66}$ Instituto de F\'{i}sica, Universidade Federal do Rio Grande do Sul (UFRGS), Porto Alegre, Brazil\\
$^{67}$ Instituto de F\'{\i}sica, Universidad Nacional Aut\'{o}noma de M\'{e}xico, Mexico City, Mexico\\
$^{68}$ iThemba LABS, National Research Foundation, Somerset West, South Africa\\
$^{69}$ Jeonbuk National University, Jeonju, Republic of Korea\\
$^{70}$ Johann-Wolfgang-Goethe Universit\"{a}t Frankfurt Institut f\"{u}r Informatik, Fachbereich Informatik und Mathematik, Frankfurt, Germany\\
$^{71}$ Korea Institute of Science and Technology Information, Daejeon, Republic of Korea\\
$^{72}$ KTO Karatay University, Konya, Turkey\\
$^{73}$ Laboratoire de Physique Subatomique et de Cosmologie, Universit\'{e} Grenoble-Alpes, CNRS-IN2P3, Grenoble, France\\
$^{74}$ Lawrence Berkeley National Laboratory, Berkeley, California, United States\\
$^{75}$ Lund University Department of Physics, Division of Particle Physics, Lund, Sweden\\
$^{76}$ Nagasaki Institute of Applied Science, Nagasaki, Japan\\
$^{77}$ Nara Women{'}s University (NWU), Nara, Japan\\
$^{78}$ National and Kapodistrian University of Athens, School of Science, Department of Physics , Athens, Greece\\
$^{79}$ National Centre for Nuclear Research, Warsaw, Poland\\
$^{80}$ National Institute of Science Education and Research, Homi Bhabha National Institute, Jatni, India\\
$^{81}$ National Nuclear Research Center, Baku, Azerbaijan\\
$^{82}$ National Research and Innovation Agency - BRIN, Jakarta, Indonesia\\
$^{83}$ Niels Bohr Institute, University of Copenhagen, Copenhagen, Denmark\\
$^{84}$ Nikhef, National institute for subatomic physics, Amsterdam, Netherlands\\
$^{85}$ Nuclear Physics Group, STFC Daresbury Laboratory, Daresbury, United Kingdom\\
$^{86}$ Nuclear Physics Institute of the Czech Academy of Sciences, Husinec-\v{R}e\v{z}, Czech Republic\\
$^{87}$ Oak Ridge National Laboratory, Oak Ridge, Tennessee, United States\\
$^{88}$ Ohio State University, Columbus, Ohio, United States\\
$^{89}$ Physics department, Faculty of science, University of Zagreb, Zagreb, Croatia\\
$^{90}$ Physics Department, Panjab University, Chandigarh, India\\
$^{91}$ Physics Department, University of Jammu, Jammu, India\\
$^{92}$ Physics Program and International Institute for Sustainability with Knotted Chiral Meta Matter (WPI-SKCM$^{2}$), Hiroshima University, Hiroshima, Japan\\
$^{93}$ Physikalisches Institut, Eberhard-Karls-Universit\"{a}t T\"{u}bingen, T\"{u}bingen, Germany\\
$^{94}$ Physikalisches Institut, Ruprecht-Karls-Universit\"{a}t Heidelberg, Heidelberg, Germany\\
$^{95}$ Physik Department, Technische Universit\"{a}t M\"{u}nchen, Munich, Germany\\
$^{96}$ Politecnico di Bari and Sezione INFN, Bari, Italy\\
$^{97}$ Research Division and ExtreMe Matter Institute EMMI, GSI Helmholtzzentrum f\"ur Schwerionenforschung GmbH, Darmstadt, Germany\\
$^{98}$ Saga University, Saga, Japan\\
$^{99}$ Saha Institute of Nuclear Physics, Homi Bhabha National Institute, Kolkata, India\\
$^{100}$ School of Physics and Astronomy, University of Birmingham, Birmingham, United Kingdom\\
$^{101}$ Secci\'{o}n F\'{\i}sica, Departamento de Ciencias, Pontificia Universidad Cat\'{o}lica del Per\'{u}, Lima, Peru\\
$^{102}$ Stefan Meyer Institut f\"{u}r Subatomare Physik (SMI), Vienna, Austria\\
$^{103}$ SUBATECH, IMT Atlantique, Nantes Universit\'{e}, CNRS-IN2P3, Nantes, France\\
$^{104}$ Sungkyunkwan University, Suwon City, Republic of Korea\\
$^{105}$ Suranaree University of Technology, Nakhon Ratchasima, Thailand\\
$^{106}$ Technical University of Ko\v{s}ice, Ko\v{s}ice, Slovak Republic\\
$^{107}$ The Henryk Niewodniczanski Institute of Nuclear Physics, Polish Academy of Sciences, Cracow, Poland\\
$^{108}$ The University of Texas at Austin, Austin, Texas, United States\\
$^{109}$ Universidad Aut\'{o}noma de Sinaloa, Culiac\'{a}n, Mexico\\
$^{110}$ Universidade de S\~{a}o Paulo (USP), S\~{a}o Paulo, Brazil\\
$^{111}$ Universidade Estadual de Campinas (UNICAMP), Campinas, Brazil\\
$^{112}$ Universidade Federal do ABC, Santo Andre, Brazil\\
$^{113}$ Universitatea Nationala de Stiinta si Tehnologie Politehnica Bucuresti, Bucharest, Romania\\
$^{114}$ University of Cape Town, Cape Town, South Africa\\
$^{115}$ University of Derby, Derby, United Kingdom\\
$^{116}$ University of Houston, Houston, Texas, United States\\
$^{117}$ University of Jyv\"{a}skyl\"{a}, Jyv\"{a}skyl\"{a}, Finland\\
$^{118}$ University of Kansas, Lawrence, Kansas, United States\\
$^{119}$ University of Liverpool, Liverpool, United Kingdom\\
$^{120}$ University of Science and Technology of China, Hefei, China\\
$^{121}$ University of South-Eastern Norway, Kongsberg, Norway\\
$^{122}$ University of Tennessee, Knoxville, Tennessee, United States\\
$^{123}$ University of the Witwatersrand, Johannesburg, South Africa\\
$^{124}$ University of Tokyo, Tokyo, Japan\\
$^{125}$ University of Tsukuba, Tsukuba, Japan\\
$^{126}$ Universit\"{a}t M\"{u}nster, Institut f\"{u}r Kernphysik, M\"{u}nster, Germany\\
$^{127}$ Universit\'{e} Clermont Auvergne, CNRS/IN2P3, LPC, Clermont-Ferrand, France\\
$^{128}$ Universit\'{e} de Lyon, CNRS/IN2P3, Institut de Physique des 2 Infinis de Lyon, Lyon, France\\
$^{129}$ Universit\'{e} de Strasbourg, CNRS, IPHC UMR 7178, F-67000 Strasbourg, France, Strasbourg, France\\
$^{130}$ Universit\'{e} Paris-Saclay, Centre d'Etudes de Saclay (CEA), IRFU, D\'{e}partment de Physique Nucl\'{e}aire (DPhN), Saclay, France\\
$^{131}$ Universit\'{e}  Paris-Saclay, CNRS/IN2P3, IJCLab, Orsay, France\\
$^{132}$ Universit\`{a} degli Studi di Foggia, Foggia, Italy\\
$^{133}$ Universit\`{a} del Piemonte Orientale, Vercelli, Italy\\
$^{134}$ Universit\`{a} di Brescia, Brescia, Italy\\
$^{135}$ Variable Energy Cyclotron Centre, Homi Bhabha National Institute, Kolkata, India\\
$^{136}$ Warsaw University of Technology, Warsaw, Poland\\
$^{137}$ Wayne State University, Detroit, Michigan, United States\\
$^{138}$ Yale University, New Haven, Connecticut, United States\\
$^{139}$ Yonsei University, Seoul, Republic of Korea\\
$^{140}$ Affiliated with an institute covered by a cooperation agreement with CERN\\
$^{141}$ Affiliated with an international laboratory covered by a cooperation agreement with CERN.\\

\end{flushleft} 

\end{document}